# Synthesis, crystal structure, and magnetic properties of $RE_3Sb_3Mg_2O_{14}$ (RE=La, Pr, Sm, Eu, Tb, Ho): new 2D Kagome materials

M.B. Sanders*, K.M. Baroudi, J.W. Krizan, O.A. Mukadam, and R.J. Cava*

*Department of Chemistry, Princeton University, Princeton, New Jersey 08544*

*Corresponding authors: marisas@princeton.edu (M.B. Sanders), rcava@princeton.edu (R.J. Cava)

**Abstract**

We present the crystal structures and magnetic properties of $RE_3Sb_3Mg_2O_{14}$ ($La_3Sb_3Mg_2O_{14}$, $Pr_3Sb_3Mg_2O_{14}$, $Sm_3Sb_3Mg_2O_{14}$, $Eu_3Sb_3Mg_2O_{14}$, $Tb_3Sb_3Mg_2O_{14}$, and $Ho_3Sb_3Mg_2O_{14}$), a family of novel materials based on a perfect geometry 2D rare earth Kagome lattice. Structure refinements were performed by the Rietveld method using X-ray diffraction data, indicating that the layered compounds are fully structurally ordered. The compounds crystallize in a rhombohedral supercell of the cubic pyrochlore structure, in the space group *R-3m*. As indicated by magnetic susceptibility measurements, they exhibit predominantly antiferromagnetic interactions between rare earth moments. Except for possibly $Pr_3Sb_3Mg_2O_{14}$ and $Eu_3Sb_3Mg_2O_{14}$, none of the compounds show any signs of magnetic ordering above 2 K. This $RE_3Sb_3Mg_2O_{14}$ family of compounds is similar to that of $RE_3Sb_3Zn_2O_{14}$, except the series reported here features a fully ordered distribution of cations in both the nonmagnetic antimony and magnesium sites and the magnetic rare earth kagome sites. The compounds appear to be relatively defect-free and are therefore model systems for investigating magnetic frustration on an ideal 2D rare earth Kagome lattice.

# 1. Introduction

With its network of corner-sharing triangles, the two-dimensional Kagome lattice is a model system for investigating geometric magnetic frustration. Two popular transition metal-based Kagome materials are Herbertsmithite, $ZnCu_3(OH)_6Cl_2$, and vesignieite, $BaCu_3V_2O_8$, which are also promising quantum spin liquid candidates.[1,2] However, there has been a dearth of information related to rare earth Kagome materials. The langasites are one example, such as $Pr_3Ga_5SiO_{14}$ and $Nd_3Ga_5SiO_{14}$, but these structures possess "breathing" Kagome planes that are characterized by alternating larger and smaller rare earth triangles.[3,4]

The 3D $A_2B_2O_7$ pyrochlore structure is another model for exploring geometric frustration. Pyrochlores with magnetic rare earth ions on the A site have been shown to exhibit exotic magnetic phenomena, such as the spin ices $Dy_2Ti_2O_7$ and $Ho_2Ti_2O_7$.[5] In the pyrochlore structure, both the A and B ions form Kagome planes along [111] directions, connected by planes of A or B triangles to create a magnetic lattice of corner sharing tetrahedra. Many experiments have been conducted to selectively dope the pyrochlore structure with nonmagnetic ions in an ordered manner in an attempt to form a $RE^{3+}$-only Kagome lattice. This would enable the study of magnetic frustration on an intrinsic 2D rare earth Kagome. Unfortunately, such attempts have often resulted in defect structures.[6]

Here we report the crystal structures and elementary magnetic properties of compounds in the $RE_3Sb_3Mg_2O_{14}$ family (RE=La, Pr, Sm, Eu, Tb, Ho), a series of new materials with a crystallographically ordered, ideal 2D Kagome lattice derived from partial ion substitution in the common cubic pyrochlore structure. This family is a variant of compounds originally reported with magnetic ions in place of the $Mg^{2+}$.[7] The structure is similar to the $A_2B_2O_7$ pyrochlore, but is two-dimensional, with rare earth ions on an ideal Kagome lattice with $MgRE_3$ in the A sites and $MgSb_3$ in the B sites. This differs from the pyrochlore, which has magnetic ions connecting its Kagome planes (Figure 1). The magnetic properties of three compounds in this new family of materials (RE=Gd, Dy, Er), which all exhibit interesting magnetic ground states, including kagome spin ice, dipolar spin order, and the KT transition, have recently been reported.[8] Just like the $RE_3Sb_3Zn_2O_{14}$ family of materials (RE=La, Pr, Nd, Sm, Eu, Gd), $RE_3Sb_3Mg_2O_{14}$ is fully layered with the rare earth ions in an ordered, symmetric Kagome array.[9] While the Zn analogs have some local disorder due to the off-center displacement of one of the $Zn^{2+}$ ions, $RE_3Sb_3Mg_2O_{14}$ has a completely ordered distribution of cations. The perfect cation array and the distinct rare earth Kagome lattice constructed by $RE^{3+}$ in $RE_3Sb_3Mg_2O_{14}$ make it an ideal model for exploring magnetic frustration in an intrinsic Kagome material. Here we describe the solid state chemistry and elementary magnetic properties of this family for a wide range of RE ions.

## 2. Experimental

Samples were synthesized by solid state reaction from $La_2O_3$, $Pr_6O_{11}$, $Sm_2O_3$, $Eu_2O_3$, $Tb_4O_7$, $Ho_2O_3$, $Sb_2O_3$, and $Mg(OH)_2$ in alumina crucibles. The rare earth oxides were first dried at 800 °C overnight. The stoichiometric mixtures were then thoroughly ground in an agate mortar and pestle and pre-reacted in air at 1000 °C for 15 hours. Following this, the mixtures were reacted in air at 1300 °C for 60 hours with repeated grindings.

The X-ray powder diffraction data were collected at room temperature using a Bruker D8 Advance Eco diffractometer with Cu Kα radiation (λ=1.5418 Å) and a Lynxeye detector. Structural refinements on $RE_3Sb_3Mg_2O_{14}$ were performed using the Rietveld method with the program Fullprof. The magnetic susceptibilities of $RE_3Sb_3Mg_2O_{14}$ were measured between 2 and 300 K in a Quantum Design Physical Properties Measurement System (PPMS) in an applied field of 5000 Oe. The magnetizations were linearly proportional to magnetic field for all temperatures above 2 K to fields of approximately 15,000 Oe in all materials; therefore, the magnetic susceptibility was defined as M/H at an intermediate field of H=5000 Oe.

## 3. Results and Discussion

*3.1 Crystal Structure of $RE_3Sb_3Mg_2O_{14}$*

The $RE_3Sb_3Mg_2O_{14}$ compounds studied here (RE=La, Pr, Sm, Eu, Tb, Ho) are all isostructural and crystallize in the rhombohedral space group *R-3m* in the hexagonal coordinate system. The structures were refined by placing Mg on the *3a* and *3b* sites and RE and Sb on the *9e* and *9d* sites. The oxygens were initially placed in the positions found in $La_3Sb_3Zn_2O_{14}$ as a guide.[9] The Rietveld refinements proceeded smoothly, the results of which are summarized in Table 1. Figures 2 through 4 show the plots of the observed and calculated X-ray diffraction profiles for the final structural models. Some selected interatomic distances and angles are listed in Table 2.

Depicted in Figure 5 are the refined crystal structure of $La_3Sb_3Mg_2O_{14}$ (a) and its individual coordination polyhedra (b). Antimony is positioned in an octahedron of oxygen atoms with bond lengths of 1.94-2.00 Å. This $SbO_6$ octahedron is reasonably normal, with O-Sb-O bond angles of 90 ± 5°. Magnesium is also in octahedral coordination with a Mg1-O3 bond length of 2.13 Å, yet with O-Mg1-O bond angles that vary from 81.3 to 98.7°. The Mg2 site is 8-coordinated by two O1 oxygens with bond lengths of 1.94 Å and a puckered 6-membered ring of O2 oxygens with Mg2-O2 distances of 2.67 Å. Lanthanum is surrounded by eight oxygens in a distorted cube with three different rare earth-oxygen bond lengths: two short La-O1 bonds (2.39 Å), two long La-O2 bonds (2.64 Å), and four intermediate La-O3 bonds (2.54 Å).

As shown in Figure 6, the lattice parameters of $RE_3Sb_3Mg_2O_{14}$ decrease with decreasing ionic radius of the $RE^{3+}$, as expected. The metal-oxygen distances are in general normal when compared to the sum of their ionic radii.[10] While the RE-O bond distances become shorter from La (2.39-2.64 Å) to Ho (2.30-2.55 Å), the Sb-O and Mg1-O bond lengths are relatively constant, consistent with expectations. For example, for $Ho_3Sb_3Mg_2O_{14}$, the Sb-O2 (1.97 Å), Sb-O3 (1.94 Å), and Mg1-O3 (2.10 Å) bond lengths correspond well with those of $La_3Sb_3Mg_2O_{14}$ (2.00 Å, 1.94 Å, 2.13 Å). The O2-Sb-O2 bond angle changes slightly but systematically in going from La (86.3°) to Ho (83.4°). The Mg2-O bond lengths also change in a systematic manner throughout the series.

$RE_3Sb_3Mg_2O_{14}$ bears a strong resemblance to its parent cubic pyrochlore when considering the role that magnesium plays in the structure's cation ordering. $Mg^{2+}$ is present to fill in the non Kagome sites in both the $RE^{3+}$ and $Sb^{5+}$ planes, to form planes of $MgRE_3$ and $MgSb_3$ on what would be the A and B sites in an $A_2B_2O_7$ pyrochlore. The crystal structure presented in Figure 5a depicts the stacking of these $MgRE_3$ and $MgSb_3$ layers in the rhombohedral-on-hexagonal cell. From this perspective, $RE_3Sb_3Mg_2O_{14}$ can also be considered as $(RE_{0.75}Mg_{0.25})_2(Sb_{0.75}Mg_{0.25})_2O_7$, with the Mg and other ions ordered; this is essentially the pyrochlore structure with an ordered substitution of ¼ $A^{3+}$ and $B^{4+}$ ions. Examination of the x-ray diffraction patterns in Figures 2-4 reveals that the strongest reflection (222) of the cubic pyrochlore (2θ~30°) has split into two separate peaks in the current compounds, even qualitatively suggesting that the structures have complete Mg-Sb and Mg-RE ordering. It is also worthwhile to point out that the RE-RE separations in $RE_3Sb_3Mg_2O_{14}$ are very similar to those observed in the parent cubic pyrochlore. For instance, the interplane Pr-Pr distance in $Pr_3Sb_3Mg_2O_{14}$ is 3.72 Å, while that in $Pr_2Zr_2O_7$ is 3.78 Å.[11]

*3.2 Magnetic Properties*

**Pr$_3$Sb$_3$Mg$_2$O$_{14}$**

The magnetic data for the compounds were fit to the Curie-Weiss law $\chi = \chi_0 + \frac{C}{T-\theta_{CW}}$ where $\chi_0$ is the temperature independent contribution to the susceptibility $\chi$, C is the Curie Constant, and $\theta_{CW}$ is the Weiss temperature. The effective moments were then obtained using $\mu_{eff} \propto 2.83\sqrt{C}$. The magnetic susceptibility and inverse susceptibility of Pr$_3$Sb$_3$Mg$_2$O$_{14}$ are shown in Figure 7. Data were fit to the Curie-Weiss law in the temperature range 150 to 300 K with a $\chi_0$ of 0.000005 emu Oe$^{-1}$ mol Pr$^{-1}$. This yielded an effective moment of 3.44 $\mu_B$ /Pr and a Weiss temperature of -33.7 K. A similar antiferromagnetic Curie-Weiss theta of -34.8 K was reported for Pr$_3$Sb$_3$Zn$_2$O$_{14}$.[9] The $\mu_{eff}$ calculated for Pr$^{3+}$ is slightly lower than the expected free ion value of 3.58 $\mu_B$. As observed in other Pr$^{3+}$-containing oxides, such as Pr$_3$Sb$_3$Zn$_2$O$_{14}$ (3.22 $\mu_B$/Pr), Pr$_3$IrO$_7$ (3.26 $\mu_B$/Pr), and Pr$_{0.01}$U$_{0.99}$O$_2$ (3.0 $\mu_B$/Pr), the observed moment of praseodymium is often reduced from the ideal due to its dependence on crystal field effects.[9,12,13] This is a direct consequence of praseodymium's excited state multiplets being located very close in energy to its ground state.[14]

The broad peak in the $\chi(T)$ plot for Pr$_3$Sb$_3$Mg$_2$O$_{14}$ at around 17 K suggests some type of magnetic ordering, but this has not yet been further investigated. The slight upswing in the susceptibility at low temperature is a consequence of the presence of either uncoupled spins or a magnetic impurity. The broad feature at around 17 K and the upturn at low temperatures were also observed in the $\chi(T)$ for Pr$_3$Sb$_3$Zn$_2$O$_{14}$.[9] The H-dependent magnetization (M) plots of Pr$_3$Sb$_3$Mg$_2$O$_{14}$ at 2 and 100 K are presented as insets to the susceptibility in Figure 7. At both high and low temperatures, the *M(H)* response is reversible and linear up to an applied field of $\mu_0$H= 9 T without any signs of saturation.

**Sm$_3$Sb$_3$Mg$_2$O$_{14}$**

As can be seen in the $\chi(T)$ plot of Sm$_3$Sb$_3$Mg$_2$O$_{14}$ in Figure 8, the high temperature susceptibility is characteristic of temperature-independent Van Vleck paramagnetism. This is common in many Sm-containing compounds, and arises from the first excited J=7/2 multiplet of Sm$^{3+}$.[15] Therefore, the Curie-Weiss law was fit at low temperature between 1.8 and 10 K, yielding a moment of 0.53 $\mu_B$/Sm and a Weiss temperature of -4.01 K. The θ$_{CW}$ indicates antiferromagnetic interactions between Sm moments. No $\chi_0$ was used in the fitting. Although the $\mu_{eff}$ is low compared to the expected free ion value of 0.83 $\mu_B$/Sm, it is similar to that of other reported $Sm^{3+}$-containing oxides, including $Sm_3Sb_3Zn_2O_{14}$ (0.53 $\mu_B$/Sm) and $Sm_2Zr_2O_7$ (0.50 $\mu_B$/Sm).[16] Samarium's large crystal field splitting of its lowest J=5/2 multiplet is likely related to the low moments observed for these compounds. The field-dependent magnetizations *M (H)* at 2 and 100 K provided in the inset to Figure 8 for $Sm_3Sb_3Mg_2O_{14}$ are both linear and reversible with the field increasing and decreasing.

**$Eu_3Sb_3Mg_2O_{14}$**

The temperature dependent magnetic susceptibility of $Eu_3Sb_3Mg_2O_{14}$ from 2 to 300 K is shown in Figure 9. As is typical for $Eu^{3+}$ compounds, $Eu_3Sb_3Mg_2O_{14}$'s susceptibility exhibits Van Vleck paramagnetism at very high temperatures, increases with decreasing temperature, and then rapidly increases at around 2-10 K.[12] Data were fit in the range 200-300 K with a $\chi_0$ of 0.001155 emu $Oe^{-1}$ mol $Eu^{-1}$, resulting in an effective moment of 3.54 $\mu_B$/Eu and $\theta_{CW}$ of -100.1 K. The Weiss temperature is indicative of strong antiferromagnetic interactions and the effective moment is close to the observed value of 3.40-3.51 $\mu_B$ / $Eu^{3+}$.[17] Similar to what was seen in the magnetic susceptibility of $Eu_3Sb_3Zn_2O_{14}$, there may be a relationship between the interaction strength and the appearance of temperature independent paramagnetism; heat capacity or neutron diffraction measurements would conclusively determine whether magnetic ordering is present. The small peak at low temperature in the susceptibility is likely due to the presence of a magnetic impurity. This slight upturn has been observed in other $Eu^{3+}$ oxides, including $Eu_3Sb_3Zn_2O_{14}$, $Eu_3MoO_7$, and $Eu_2Ti_2O_7$.[9,12,18] The magnetization at various applied fields for $Eu_3Sb_3Mg_2O_{14}$ is shown in the inset to Figure 9. At both 2 and 100 K the *M(H)* plots are linear and reversible up to $\mu_0 H$= 9 T.

**$Tb_3Sb_3Mg_2O_{14}$**

The temperature dependence of magnetic susceptibility of $Tb_3Sb_3Mg_2O_{14}$ is shown in Figure 10. $\chi(T)$ was fit to the Curie-Weiss law from 150 to 300 K with a $\chi_0$ of 0.002 emu $Oe^{-1}$ mol $Tb^{-1}$. An effective magnetic moment of 9.67 $\mu_B$/Tb was obtained from the Curie-Weiss fit, which is close to the expected free ion value of 9.70 $\mu_B$/Tb. A Weiss temperature of -8.86 K was extrapolated from the high temperature fit, which suggests the presence of antiferromagnetic interactions between $Tb^{3+}$ spins. The field dependent magnetization plots of $Tb_3Sb_3Mg_2O_{14}$ at 2 and 100 K are displayed as insets to the magnetic susceptibility in Figure 10. At 100 K, the *M(H)* plot is linear and reversible up to an applied field of $\mu_0H$= 9 T. However, at 2 K, the response is nonlinear with the onset of saturation at roughly 5.8 $\mu_B$/Tb.

**$Ho_3Sb_3Mg_2O_{14}$**

Shown in Figure 11 are the magnetic susceptibility and inverse susceptibility for $Ho_3Sb_3Mg_2O_{14}$. Data were fit to the Curie Weiss law between 100 and 300 K with a $\chi_0$ of 0.0027 emu $Oe^{-1}$ mol $Ho^{-1}$. This yielded a moment of 10.60 $\mu_B$/Ho and a Weiss temperature of -13.7 K. The effective moment matches well with the expected free ion value of 10.60 $\mu_B$/Ho. The $\theta_{CW}$ is indicative of antiferromagnetic interactions. The *M(H)* plots at 2 and 100 K are shown in the inset to Figure 11. The field dependent magnetization at 100 K is linear and reversible with the field increasing and decreasing. Conversely, the *M(H)* plot at 2 K reveals a nonlinear response between magnetization and the applied field and saturation at roughly 5.6 $\mu_B$/Ho, a little more than half the value of the effective moment of $Ho^{3+}$ obtained from the Curie-Weiss fits. A saturation value of about half the effective moment in polycrystalline samples is typically attributed to powder averaging of Ising spins; this phenomenon has been observed in several holmium pyrochlores.[19] Where overlapping, the results obtained here for all materials are in good agreement with the results reported by Dun *et. al*.

## 4. Conclusions

We have synthesized several new compounds in the $RE_3Sb_3Mg_2O_{14}$ family (RE=La, Pr, Sm, Eu, Tb, Ho). Structural refinements indicate they are fully ordered and layered and crystallize in the space group *R-3m* derived from the cubic pyrochlore structure. Magnetic property measurements show that the compounds display antiferromagnetic interactions, and no signs of magnetic ordering above 2 K except possibly the Pr and Eu cases. It is worthwhile to note that this structure type exhibits a wide range of chemical tolerance; $Mg^{2+}$ has already been substituted for $Zn^{2+}$, $Co^{2+}$, $Mn^{2+}$, and $Ca^{2+}$.[9,7,20,21] The 3D pyrochlore analogs of these compounds have been shown to exhibit several exotic magnetic phenomena at low temperatures—from the spin ice behavior observed in $Ho_2Ti_2O_7$ and $Ho_2Sn_2O_7$, to the spin liquid phenomena reported in

Tb$_2$Ti$_2$O$_7$.[22,23,5] Further characterization of these new Kagome compounds, especially at lower temperatures, could also potentially lead to novel and interesting magnetic phenomena.

**Acknowledgements**

All of this research was supported by the U.S. Department of Energy, Division of Basic Energy Sciences, Grant No. DE-FG02-08ER46544, through the Institute for Quantum Matter at Johns Hopkins University.

## Tables

**Table 1: Refinement Summary of $RE_3Sb_3Mg_2O_{14}$ @ 298 K**

|  | La | Pr | Sm | Eu | Tb | Ho |
|---|---|---|---|---|---|---|
| **RE *(9e)*** | (½,0,0) | (½,0,0) | (½,0,0) | (½,0,0) | (½,0,0) | (½,0,0) |
| **B ($Å^2$)** | 2.64(4) | 3.85(3) | 3.64(3) | 3.44(9) | 2.87(2) | 3.89(5) |
| **Mg1 *(3a)*** | (0,0,0) | (0,0,0) | (0,0,0) | (0,0,0) | (0,0,0) | (0,0,0) |
| **B ($Å^2$)** | 2.00(8) | 2.10(9) | 2.89(2) | 2.49(2) | 1.00(1) | 1.55(9) |
| **Mg2 *(18g)*** | (0,0,½) | (0,0,½) | (0,0,½) | (0,0,½) | (0,0,½) | (0,0,½) |
| **B ($Å^2$)** | 3.15(7) | 2.74(3) | 2.66(8) | 2.21(8) | 1.43(1) | 1.12(9) |
| **Sb *(9d)*** | (1/2,0,1/2) | (1/2,0,1/2) | (1/2,0,1/2) | (1/2,0,1/2) | (1/2,0,1/2) | (1/2,0,1/2) |
| **B ($Å^2$)** | 2.53(2) | 3.56(8) | 3.91(6) | 3.08(6) | 2.45(1) | 2.90(7) |
| **O1 *(6c)*** | (0,0,z) | (0,0,z) | (0,0,z) | (0,0,z) | (0,0,z) | (0,0,z) |
| *z* | 0.3903(1) | 0.3878(7) | 0.3873(4) | 0.3866(1) | 0.3866(1) | 0.3872(8) |
| **B ($Å^2$)** | 2.32(4) | 2.26(1) | 2.64(1) | 3.60(1) | 2.71(4) | 2.90(8) |
| **O2 *(18h)*** | (x,x̄,z) | (x,x̄,z) | (x,x̄,z) | (x,x̄,z) | (x,x̄,z) | (x,x̄,z) |
| *x* | 0.5370(1) | 0.5361(4) | 0.5352(3) | 0.5346(8) | 0.5324(1) | 0.5324(9) |
| *z* | 0.1471(9) | 0.1464(9) | 0.1462(6) | 0.1459(6) | 0.1456(1) | 0.1459(2) |
| **B ($Å^2$)** | 3.39(2) | 3.29(8) | 2.31(2) | 3.26(7) | 2.19(2) | 2.60(8) |
| **O3 *(18h)*** | (x,x̄,z) | (x,x̄,z) | (x,x̄,z) | (x,x̄,z) | (x,x̄,z) | (x,x̄,z) |
| *x* | 0.1439(1) | 0.1451(6) | 0.1466(1) | 0.1471(1) | 0.1475(9) | 0.1477(1) |
| *z* | -0.0581(7) | -0.0569(8) | -0.0559(1) | -0.0557(3) | -0.0545(4) | -0.0550(2) |
| **B ($Å^2$)** | 2.52(8) | 3.96(9) | 2.29(6) | 3.61(7) | 2.12(5) | 2.76(9) |
| ***a* (Å)** | 7.50202(2) | 7.43870(7) | 7.38615(1) | 7.37237(2) | 7.33503(5) | 7.30440(9) |
| ***c* (Å)** | 17.66983(2) | 17.58285(1) | 17.47037(1) | 17.42688(9) | 17.32900(6) | 17.25801(8) |
| $R_{wp}$ *(%)* | 12.9 | 17.6 | 18.2 | 15.2 | 19.1 | 19.0 |
| $R_p$ *(%)* | 11.4 | 16.9 | 18.5 | 15.8 | 17.7 | 17.7 |
| $\chi^2$ | 2.79 | 3.20 | 3.05 | 2.25 | 3.07 | 3.85 |

**Table 2: Selected bond distanced (Å) and bond angles (°) for $RE_3Sb_3Mg_2O_{14}$ (RE = La, Pr, Sm, Eu, Tb, Ho @ 298 K)**

| RE | La | Pr | Sm | Eu | Tb | Ho |
|---|---|---|---|---|---|---|
| **$REO_8$** | | | | | | |
| **RE-O1** | 2.388(2) x2 | 2.351(7) ×2 | 2.331(6) ×2 | 2.321(9) ×2 | 2.309(9) ×2 | 2.304(9) ×2 |
| **RE-O2** | 2.644(8) x2 | 2.617(3) ×2 | 2.594(6) ×2 | 2.581(8) ×2 | 2.556(3) ×2 | 2.551(5) ×2 |
| **RE-O3** | 2.544(5) x2 | 2.507(9) ×2 | 2.472(6) ×2 | 2.463(7) ×2 | 2.439(1) ×2 | 2.431(4) ×2 |
| | 2.545(1) x2 | 2.507(4) ×2 | 2.473(1) ×2 | 2.463(1) ×2 | 2.439(6) ×2 | 2.431(9) ×2 |
| **Average RE-O** | 2.530(7) | 2.496(1) | 2.468(1) | 2.457(6) | 2.436(3) | 2.430(1) |
| **Intraplane RE-RE** | 3.751(1) | 3.719(3) | 3.693(1) | 3.686(1) | 3.667(5) | 3.652(2) |
| **Interplane RE-RE** | 5.889(9) | 5.860(9) | 5.823(4) | 5.808(9) | 5.776(3) | 5.752(7) |
| **$SbO_6$** | | | | | | |
| **Sb-O2** | 2.00(1)×4 | 1.98(8) ×4 | 1.97(9) ×4 | 1.97(8) ×4 | 1.97(9) ×4 | 1.96(9) ×4 |
| **Sb-O3** | 1.93(9)×2 | 1.94(8)×2 | 1.95(2) ×2 | 1.94(9) ×2 | 1.95(8) ×2 | 1.94(2) ×2 |
| **Average Sb-O** | 1.97(9) | 1.97(5) | 1.97(1) | 1.96(9) | 1.97(2) | 1.96(1) |
| **O2-Sb-O3** | 86.0(5) ×4 | 85.2(5) ×4 | 84.6(9) ×4 | 84.4(2) ×4 | 84.0(5) ×4 | 84.1(9) ×4 |
| | 93.9(5) ×4 | 94.7(5) ×4 | 95.3(2) ×4 | 95.5(8) ×4 | 95.9(5) ×4 | 95.8(1) ×4 |
| **O2-Sb-O2** | 93.7(5) ×2 | 94.2(2) ×2 | 94.8(1) ×2 | 95.1(3) ×2 | 96.6(3) ×2 | 96.6(4) ×2 |
| | 86.2(5) ×2 | 85.7(8) ×2 | 85.1(9) ×2 | 84.8(7) ×2 | 83.3(7) ×2 | 83.3(6) ×2 |
| **$Mg1O_6$** | | | | | | |
| **Mg1-O3** | 2.13(3) ×6 | 2.12(1) ×6 | 2.11(4) ×6 | 2.11(4) ×6 | 2.09(9) ×6 | 2.09(5) ×6 |
| **$Mg2O_8$** | | | | | | |
| **Mg2-O1** | 1.93(8) × 2 | 1.97(2) ×2 | 1.96(8) ×2 | 1.97(6) ×2 | 1.96(5) ×2 | 1.94(5) ×2 |
| **Mg2-O2** | 2.66(8)×6 | 2.63(6)×6 | 2.60(7) ×6 | 2.59(6) ×6 | 2.55(5) ×6 | 2.54(4) ×6 |
| **Average Mg2-O** | 2.48(6) | 2.47(1) | 2.44(7) | 2.44(1) | 2.40(7) | 2.39(4) |

## Figure Captions

**Figure 1**

Ball and stick figures showing the connectivity of the rare earth atoms in the $A_2B_2O_7$ pyrochlore (left) and in the $RE_3Sb_3Mg_2O_{14}$ Kagome (right) along the [111] and [001] directions, respectively. The rare earth ions in $RE_3Sb_3Mg_2O_{14}$ make up discrete Kagome layers, while in $A_2B_2O_7$, a triangular layer of magnetic ions is positioned between the Kagome planes. The cubic and rhombohedral unit cells are drawn in the background of each model.

**Figure 2**

Rietveld refinement of $La_3Sb_3Mg_2O_{14}$ (upper panel) and $Pr_3Sb_3Mg_2O_{14}$ (lower panel) at room temperature using x-ray diffraction data. The experimental pattern is in red, the calculated pattern in black, and the difference plot in blue, showing the very good fit of the model to the data. The green marks indicate Bragg reflections.

**Figure 3**

Rietveld refinement of $Sm_3Sb_3Mg_2O_{14}$ (upper panel) and $Eu_3Sb_3Mg_2O_{14}$ (lower panel) at room temperature using x-ray diffraction data. The experimental pattern is in red, the calculated pattern in black, and the difference plot in blue, showing the very good fit of the model to the data. The green marks indicate Bragg reflections.

**Figure 4**

Rietveld refinement of $Tb_3Sb_3Mg_2O_{14}$ (upper panel) and $Ho_3Sb_3Mg_2O_{14}$ (lower panel) at room temperature using x-ray diffraction data. The experimental pattern is in red, the calculated pattern in black, and the difference plot in blue, showing the very good fit of the model to the data. The green marks indicate Bragg reflections.

**Figure 5**

A. The crystal structure of $La_3Sb_3Mg_2O_{14}$, depicting the stacking of $MgLa_3$ and $MgSb_3$ layers. The violet polyhedra represent $LaO_8$, the green polyhedra $SbO_6$, and the two pink polyhedra $Mg1O_6$ and $Mg2O_8$, respectively. B. The coordination polyhedra of

La$_3$Sb$_3$Mg$_2$O$_{14}$: La (upper left), Sb (lower left), Mg1 (upper right), and Mg2 (lower right). The different types of oxygen atoms are represented in different shades of blue.

**Figure 6**

The *a* and *c* lattice parameters for RE$_3$Sb$_3$Mg$_2$O$_{14}$ as a function of rare earth ionic radius.[10] The purple and pink circles represent the *a* and *c* parameters, respectively. The standard deviations are smaller than the plotted points and so error bars are excluded from the figure.

**Figure 7**

The DC magnetic susceptibility and inverse susceptibility of Pr$_3$Sb$_3$Mg$_2$O$_{14}$ measured in an applied field of 5000 Oe. The Curie-Weiss fit is shown in black. The insets to the right of the MT plot show the magnetization as a function of applied field *M(H)* at 2 K (upper panel) and 100 K (lower panel).

**Figure 8**

Temperature-dependent magnetic susceptibility of Sm$_3$Sb$_3$Mg$_2$O$_{14}$ measured in an applied field of 5000 Oe. The low temperature Curie-Weiss fit is shown in cyan. The insets to the right of the magnetic susceptibility plot display the field-dependent magnetizations at 2K and 100 K.

**Figure 9**

The DC magnetic susceptibility and reciprocal susceptibility of Eu$_3$Sb$_3$Mg$_2$O$_{14}$ measured in an applied field of 5000 Oe. The Curie-Weiss fit is shown in black. The plots in the upper and lower panels to the right of the MT graph show the field-dependent magnetizations at 2 and 100 K, respectively.

**Figure 10**

Temperature-dependent magnetic susceptibility of Tb$_3$Sb$_3$Mg$_2$O$_{14}$ measured in an applied field of 5000 Oe. The Curie-Weiss fit is shown in black. The insets display *M(H)* plots of the compound at temperatures of 2 K (upper panel) and 100 K (lower panel).

**Figure 11**

Magnetic susceptibility and inverse susceptibility of $Ho_3Sb_3Mg_2O_{14}$ measured in an applied field of 5000 Oe. The Curie-Weiss fit is shown in light green. The panels to the right of the susceptibility plot show the field-dependent magnetizations at temperatures of 2 and 100 K.

**Figures**
**Figure 1**

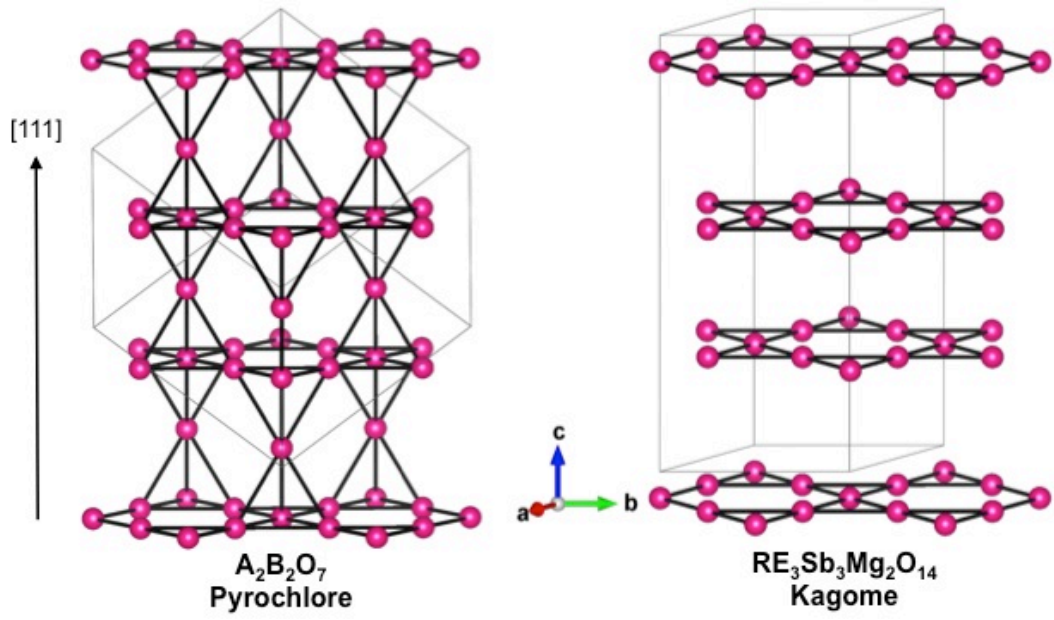

**Figure 2**

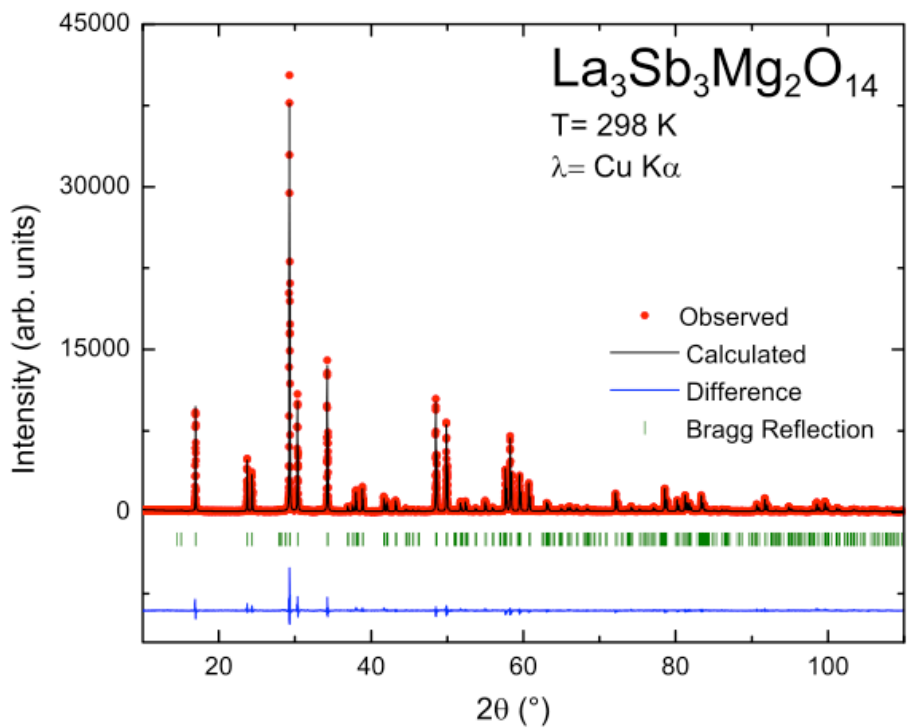

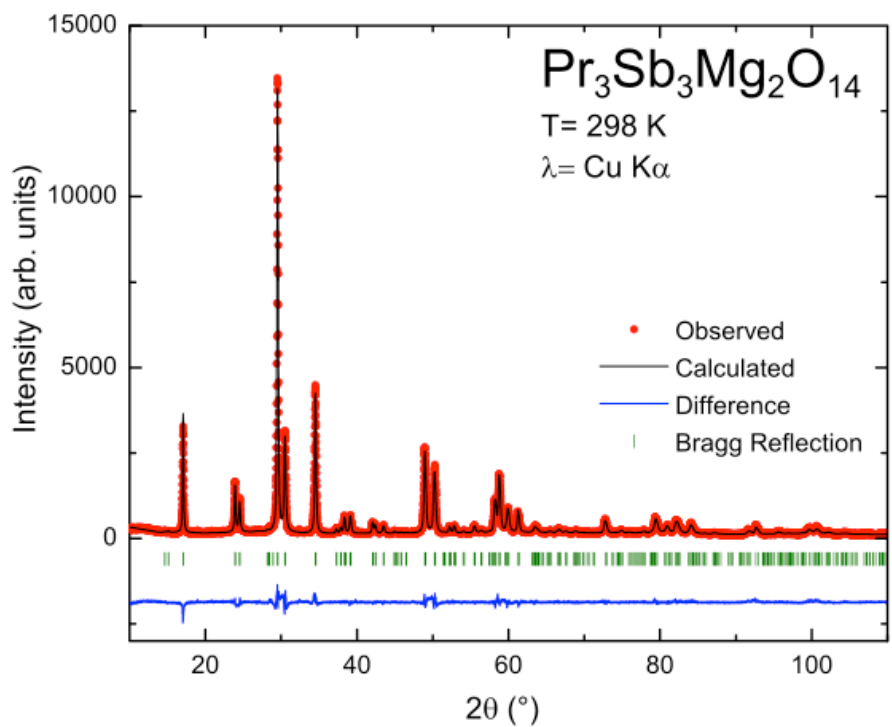

**Figure 3**

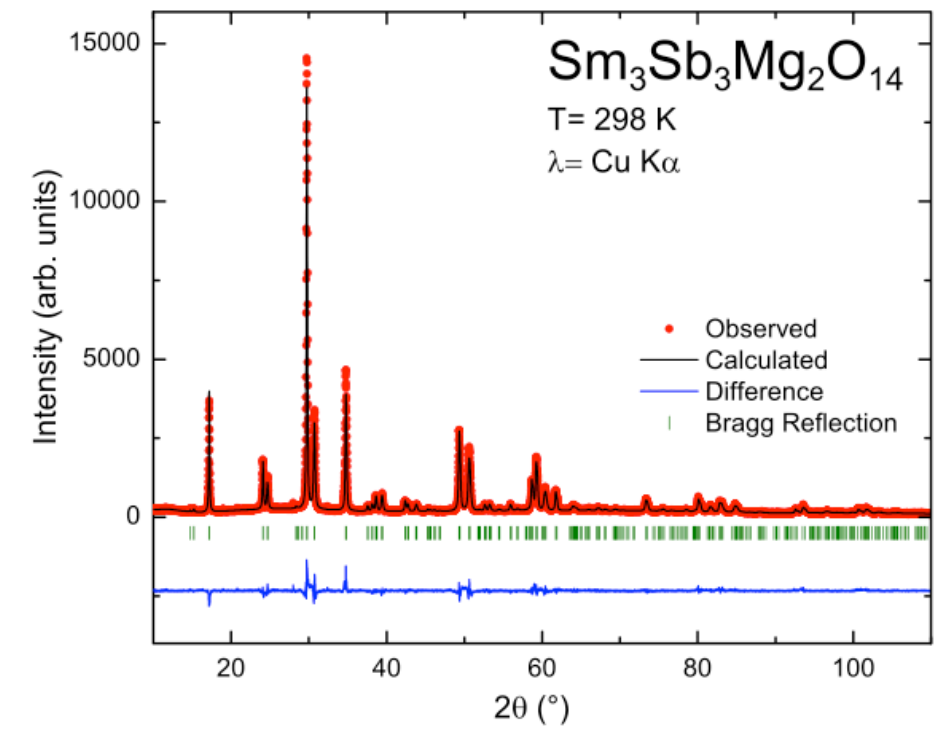

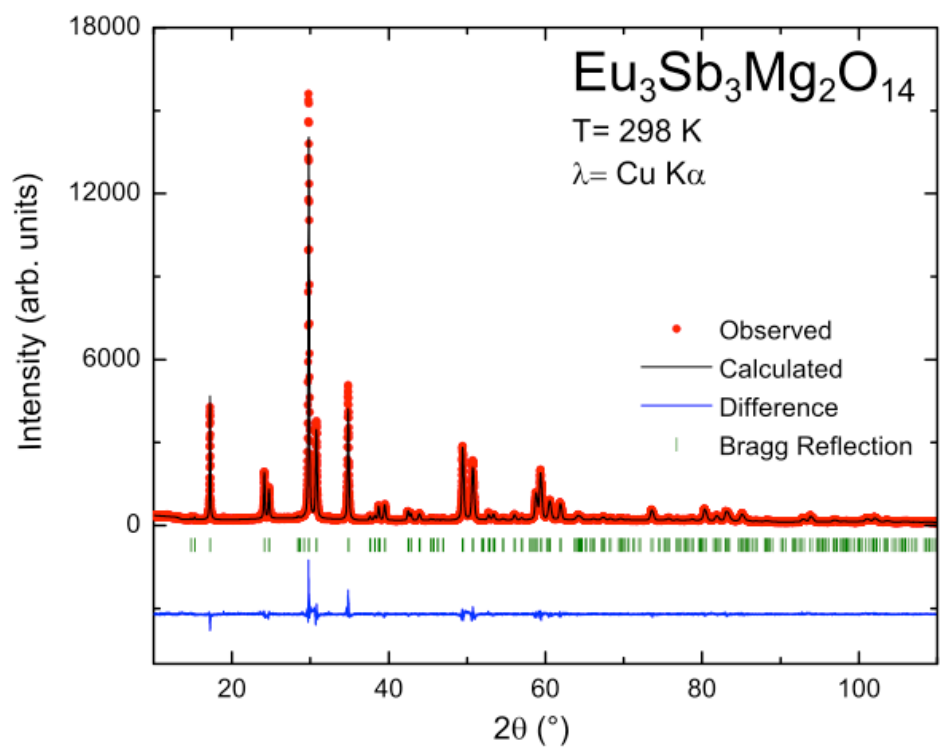

**Figure 4**

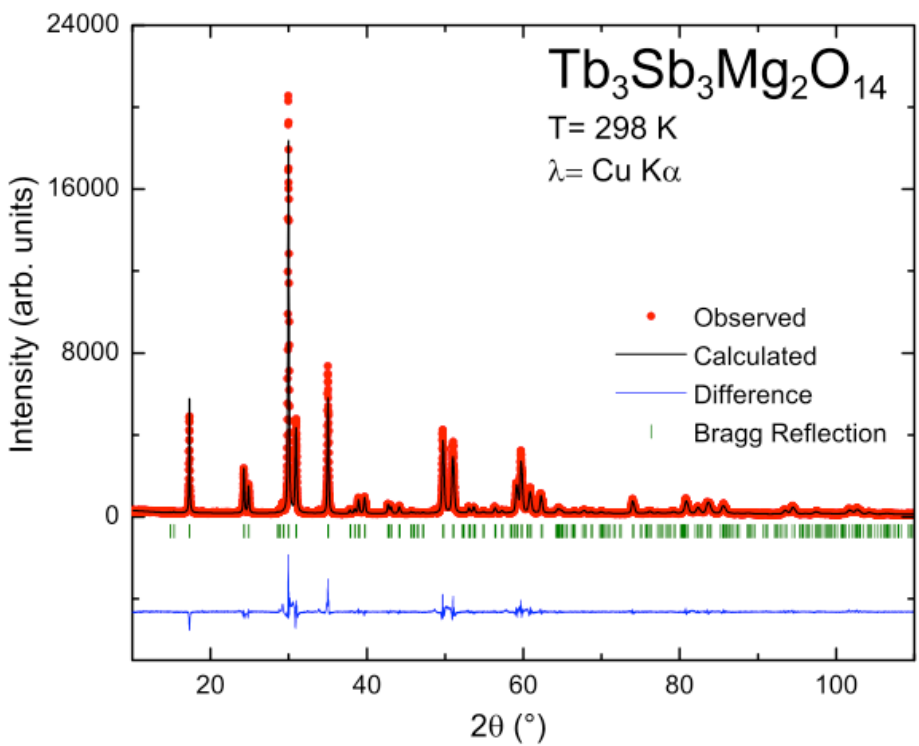

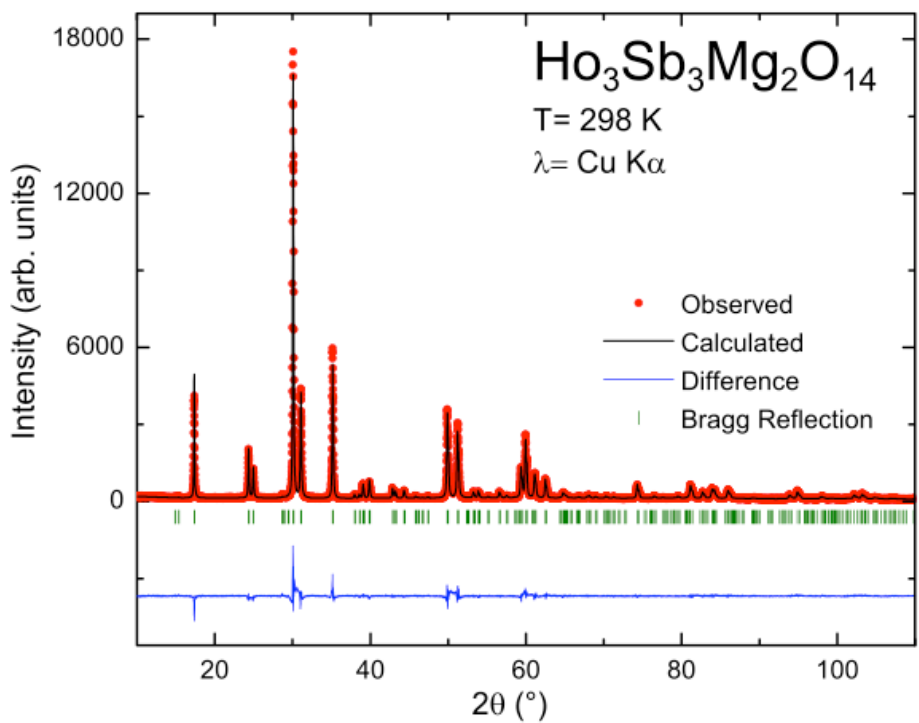

**Figure 5**

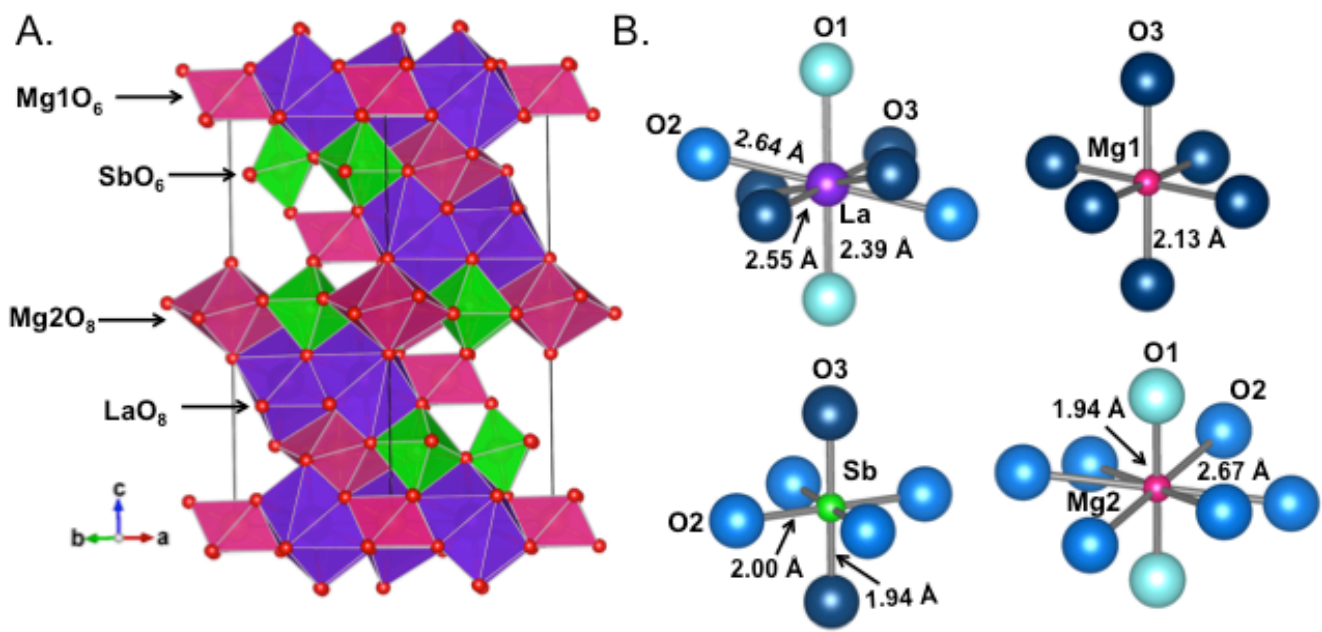

**Figure 6**

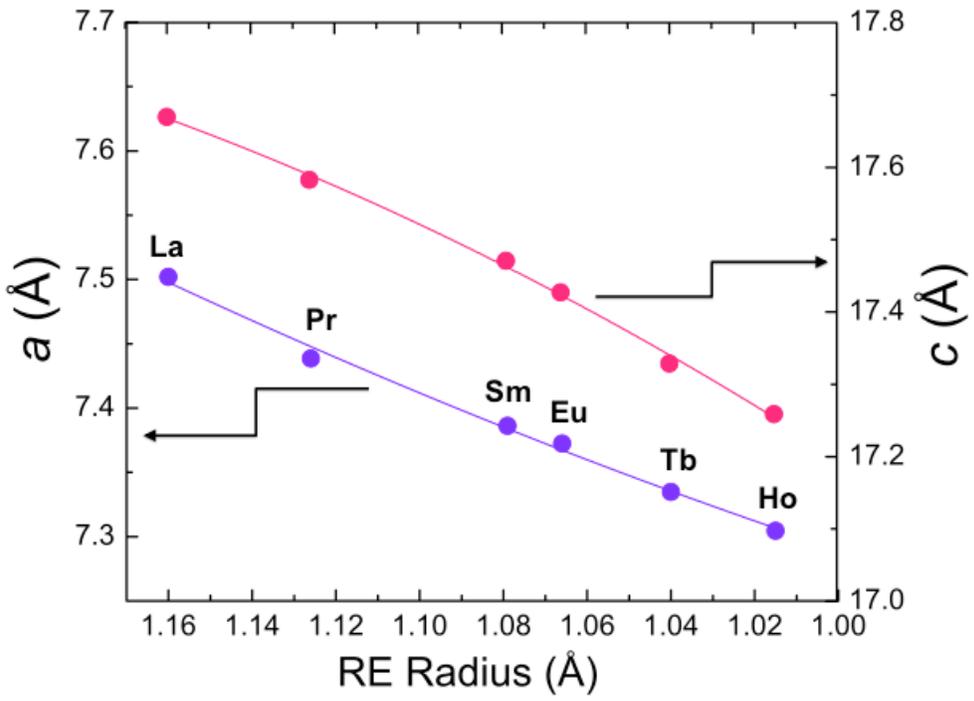

**Figure 7**

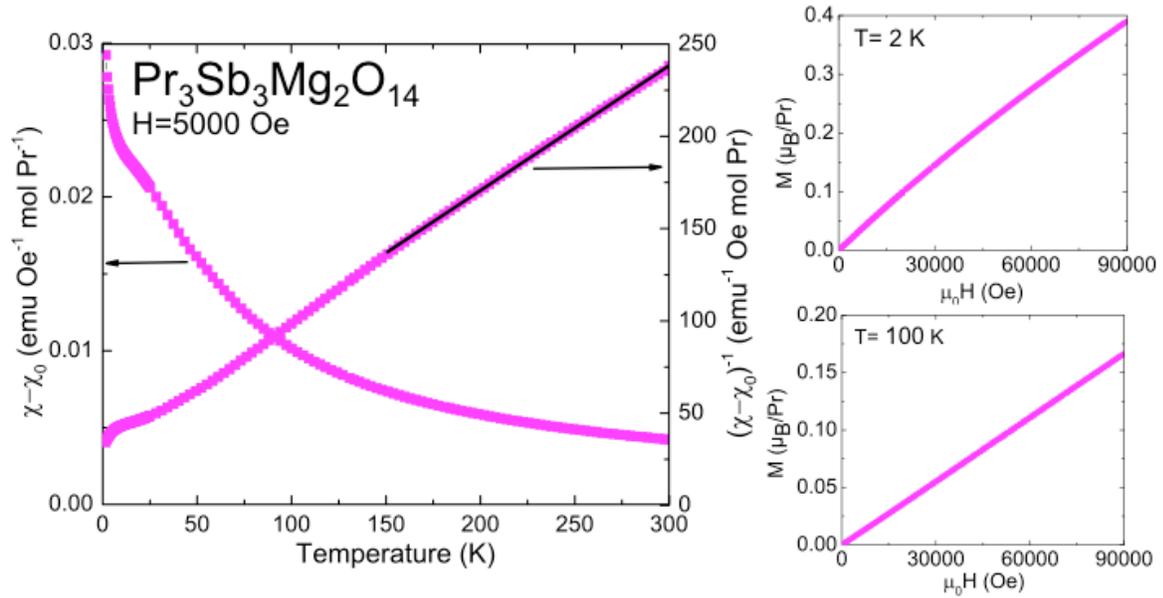

**Figure 8**

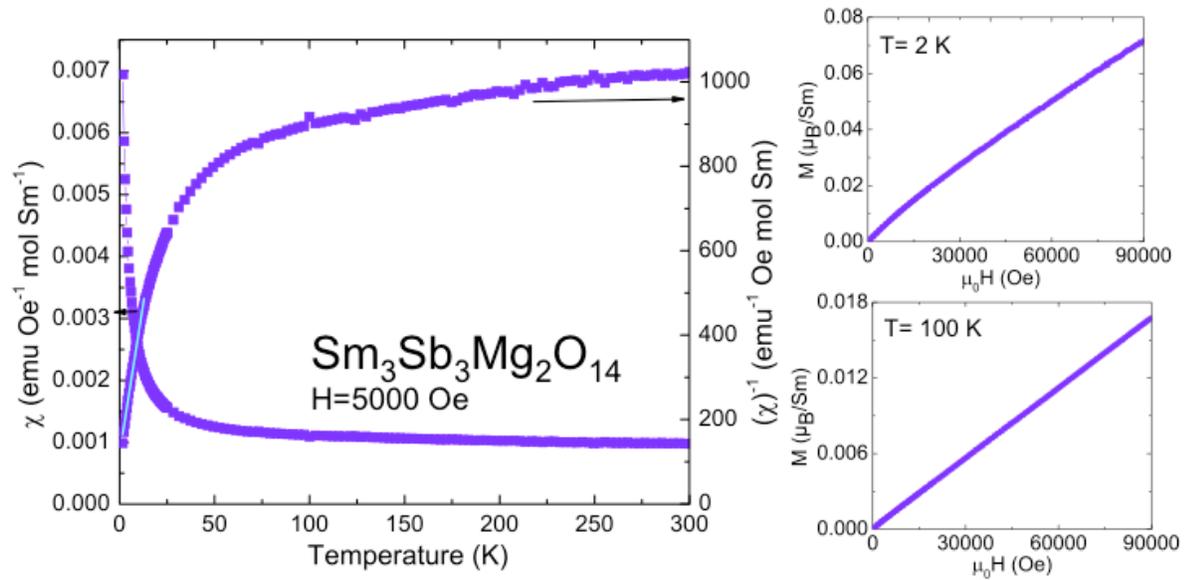

**Figure 9**

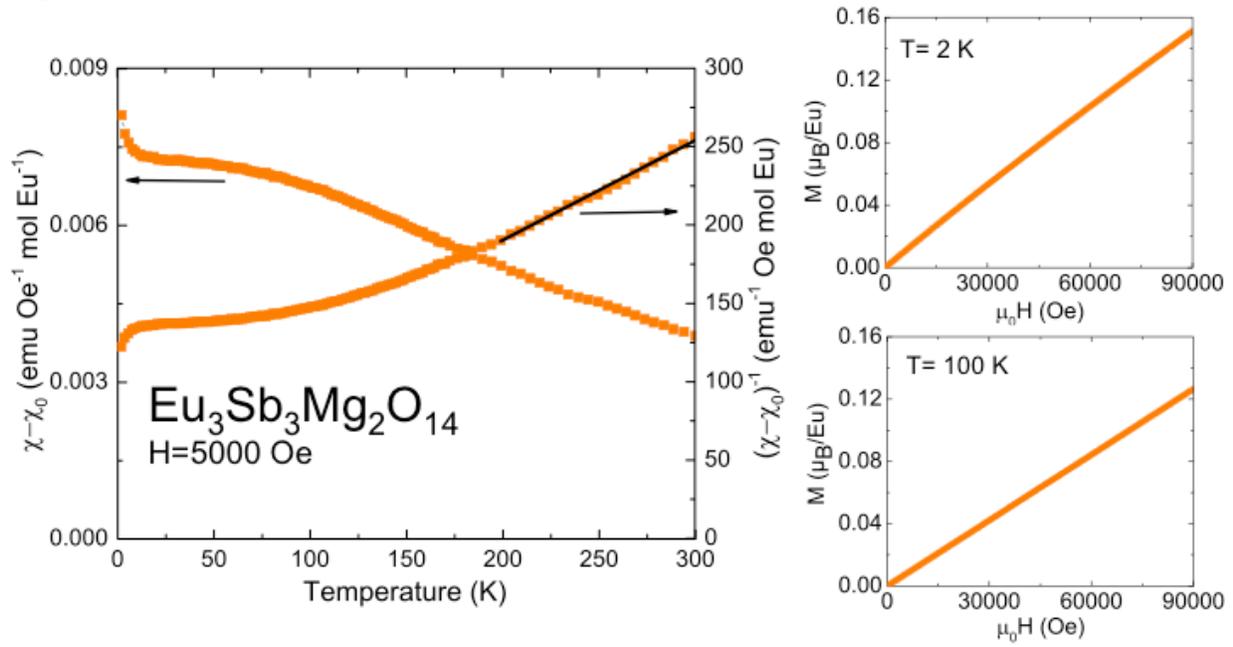

**Figure 10**

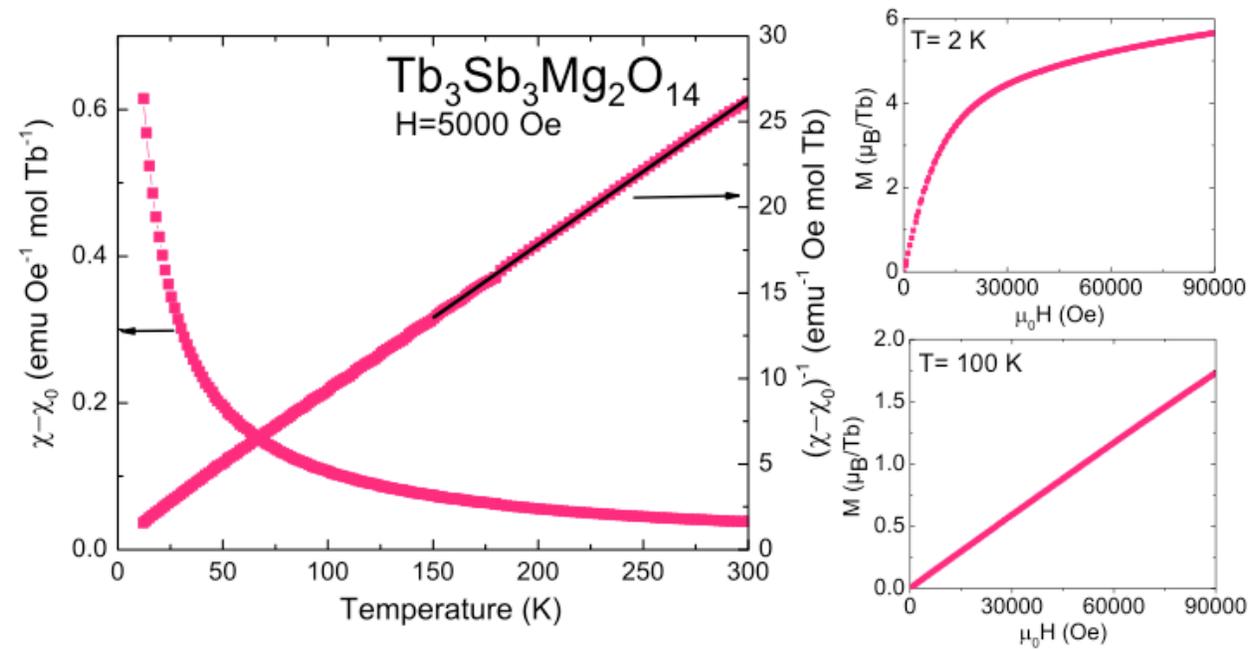

**Figure 11**

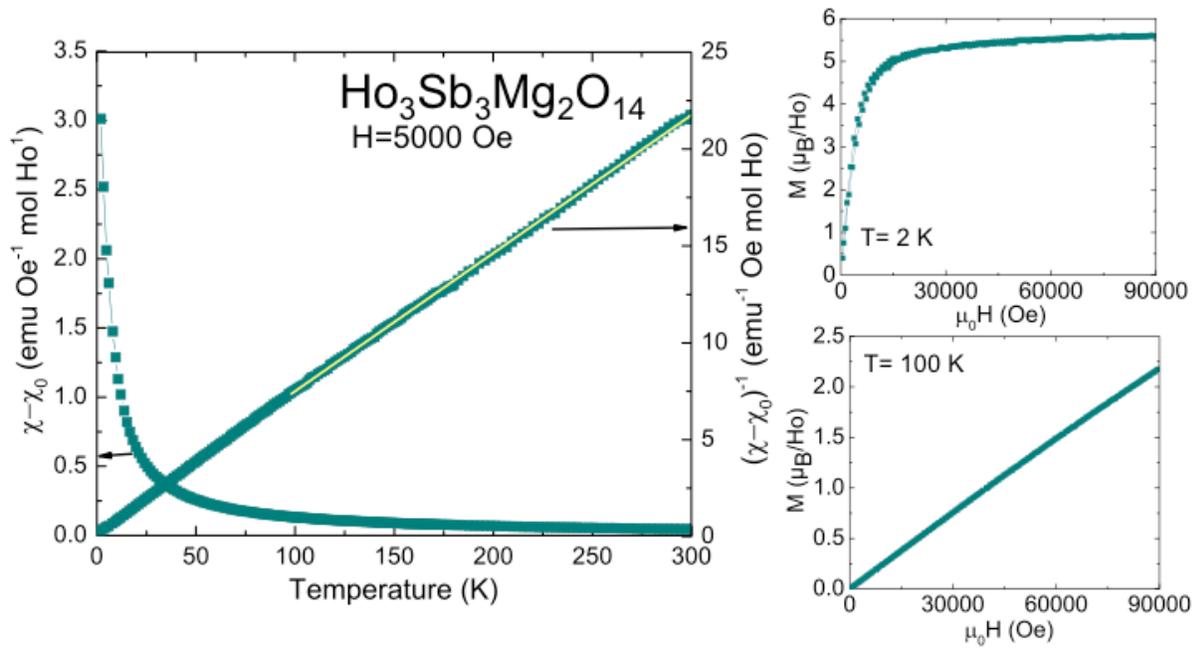